\documentclass[runningheads]{llncs}
\usepackage{graphicx,xcolor}
\usepackage{cite}
\begin{document}
\title{How Correlated are Community-aware and Classical Centrality Measures in Complex Networks?}
%
%
\author{Stephany Rajeh* \orcidID{0000-0002-7686-8506} \and
Marinette Savonnet \orcidID{0000-0003-0449-5277} \and
Eric Leclercq \orcidID{0000-0001-6382-2288} \and
Hocine Cherifi \orcidID{0000-0001-9124-4921}}
\authorrunning{S. Rajeh et al.}
\titlerunning{Correlation of Centrality Measures}
%
\institute{Laboratoire d’Informatique de Bourgogne - University of Burgundy, Dijon, France
\email{*stephany.rajeh@u-bourgogne.fr}}
\maketitle       
%
\begin{abstract}
Unlike classical centrality measures, recently developed community-aware centrality measures use a network's community structure to identify influential nodes in complex networks. This paper investigates their relationship on a set of fifty real-world networks originating from various domains. Results show that classical and community-aware centrality measures generally exhibit low to medium correlation values. These results are consistent across networks. Transitivity and efficiency are the most influential macroscopic network features driving the correlation variation between classical and community-aware centrality measures. Additionally, the mixing parameter, the modularity, and the Max-ODF are the main mesoscopic topological properties exerting the most substantial effect.

\end{abstract}

\section*{Introduction}
Identifying influential nodes is crucial for accelerating or mitigating propagation processes in complex networks. To this end, numerous classical centrality measures relying on various topological properties have been proposed. One can distinguish two main categories: local and global measures \cite{lu2016vital}. Local metrics use information in the node neighborhood while global ones gather information from the whole network. Note that some works combine local and global information \cite{ibnoulouafi2018m}. 

Another set of centrality measures uses information on the community structure to quantify the influence of the nodes. In this paper, we refer to them as ``community-aware'' centrality measures. Unlike classical centrality measures, community-aware centrality measures distinguish intra-community links from inter-community links. Intra-community links join nodes from the same community. They are related to the node's local influence inside its community. Inter-community links join nodes belonging to different communities. Therefore, they quantify the node's impact at the global level.
Community-aware centrality measures differ based on how they integrate the intra-community and inter-community links. Community Hub-Bridge proposed by \cite{ghalmane2019immunization} selects hubs within large communities and bridges simultaneously. Comm centrality \cite{gupta2016centrality} combines the intra-community and inter-community links of a node by prioritizing the latter. Community-based Centrality \cite{zhao2015community} weights an intra-community link by its community size and an inter-community link by the size of the communities it is joining. K-shell with Community \cite{luo2016identifying} is based on the linear combination of the k-shell of a node by considering the intra-community links and inter-community links networks separately. Participation Coefficient \cite{guimera2005functional} and Community-based Mediator \cite{tulu2018identifying} tends to select important nodes based on the heterogeneity of their intra-community and inter-community links. The Participation Coefficient of a node decreases if it doesn't participate in any other community than its own. Community-based Mediator reduces to the normalized degree centrality if the proportion of intra-community and inter-community links of a node are equal. Modularity Vitality \cite{modvitality} is a signed community-aware centrality measure. It is based on the modularity variation when removing a node in the network. Since bridges connect different communities, their presence decreases modularity. Therefore, nodes with negative Modularity Vitality values are bridges. In contrast, since hubs tend to increase a network's modularity, nodes with positive Modularity Vitality values are local hubs.
Many studies are devoted to the interactions between classical centrality measures \cite{li2015correlation, ronqui2015analyzing, schoch2017correlations, rajeh2020interplay, oldham2019consistency}. However, the relationship between classical and community-aware centrality measures is almost unexplored \cite{rajeh2020investigating}. Our goal in this paper is to gain a better understanding of this issue. In other words, we intend to answer the following questions: \newline
1) What is the relationship between classical and community-aware centrality measures? \newline
2) What is the influence of the macroscopic and mesoscopic topological properties on their relationship?

The paper is organized as follows. First, the classical and community-aware centrality measures are introduced. In the subsequent two sections, the analyses of the correlation and the network topology are presented. Finally, the conclusion is given.

\section*{Classical and Community-aware Centrality measures}

This study investigates ten classical centrality measures, of which five are local (Degree, Leverage, Laplacian, Diffusion Degree, and Maximum Neighborhood Component) and five are global (Betweenness, Closeness, Katz, PageRank, and Subgraph). Table \ref{tab_classical_centrality}
reports their definition. They are compared with seven community-aware measures introduced earlier and described in table \ref{tab_communityaware_centrality}. Table \ref{tab_data} quotes the fifty real-world networks used in the experiments. They are from various domains (animal, biological, collaboration, online/offline social networks, infrastructural, and miscellaneous). Since the community structure is sensitive to the community detection algorithm, Louvain and Infomap \cite{orman2011qualitative} are used to extract intra-community and inter-community links. Due to space constraints, the networks' topological characteristics and results based on Louvain are provided in the supplementary materials\footnote{https://github.com/StephanyRajeh/MixedCommunityAwareCentralityAnalysis}. Furthermore, as there are no fundamental differences, we restrict our attention in analyzing the results based on the community structure revealed using Infomap.

\begin{table}[ht!]
 \begin{center}
   \begin{tabular}{p{6cm}|p{6cm}}
   \hline\hline
   Centrality measure description & Definition \\
   \hline
   \textbf{Degree}: based on the total sum of the \newline neighbors of a node &$
\alpha_{d}(i)=\sum_{j=1}^{N}a_{ij}
$    \\

   \textbf{Leverage}: a signed centrality based on the quantity of connections compared to its neighbors &   $
\alpha_{lev}(i)=\frac{1}{k_i}\sum_{j=1}^{N}\frac{k_i - k_j}{k_i + k_j}
$  \\
   
    \textbf{Laplacian}: based on how much damage a node causes in the network after its removal
 &   $
\alpha_{lap}(i)= k_i^2 + k_i + 2\sum_{j \in \mathcal{N}_1(i)}k_j
$   \\

    \textbf{Diffusion}: based on the diffusive power of a node and that of its neighbors weighted by their propagation probabilities &   $
\alpha_{dif}(i)= \varpi_i \times \alpha_{d}(i) + \sum_{j \in \mathcal{N}_1(i)} \varpi_j \times \alpha_{d}(j) 
$  \\
   
     \textbf{Maximum Neighbor. Component}: based on the node's largest connected component (LCC) size established by its neighborhood & $
\alpha_m(i)= |LCC \in \mathcal{N}_1(i)|
$    \\

   \textbf{Betweenness}: based on the number of \newline shortest path a node falls in between\newline two other nodes &   $
\alpha_b(i)=\sum_{s,t\neq i }{\frac{\sigma_{i}(s,t)}{\sigma(s,t)} }
$    \\

    \textbf{Closeness}: based on how close, on average, a node is to all other nodes in the network &  $
\alpha_c(i)=\frac{N-1}{\sum_{j=1}^{N-1}d(i,j)}
$

   \\

   \textbf{Katz}: based on the quantity, quality, and the subsequent distances of other nodes connected to a specific node &    $
\alpha_k(i)= \sum_{p=1} \sum_{j=1} s^p a^p_{ij}
$
     \\

   \textbf{PageRank}: based on the quantity and quality of nodes connected to a specific node under a Markov chain process &  $
\alpha_p(i)=\frac{1-d}{N} + d \sum_{j \in \mathcal{N}_1(i)} \frac{\alpha_p(j)}{k_j}
$   \\

    \textbf{Subgraph}: based on a node's participation in closed walks, with paths starting and ending with the same node &   $
\alpha_s(i) = \sum_{j=1}^{N}(v_j^{i})^2 e^{\lambda_{j}}
$   \\

   \hline
   \end{tabular}
\caption{Definitions of classical centrality measures ($\alpha(i)$). $a_{i,j}$ denotes the connectivity of a node $i$ to node $j$ from the adjacency matrix $A$. $N$ is the total number of nodes.  $k_i$ and $k_j$ are the degrees of nodes $i$ and $j$, respectively. $\mathcal{N}_1(i)$ is the set of direct neighbors of node $i$. $\varpi_i$ and $\varpi_j$ are the propagation probabilities of nodes $i$ and nodes $j$, respectively ($\varpi$  is set to 1 for all nodes in this study). $\sigma(s,t)$ is the number of shortest paths between nodes $s$ and $t$
and $\sigma_{i}(s,t)$ is the number of shortest paths between nodes $s$ and $t$ that pass through node $i$. $d(i,j)$ is the shortest-path distance between node $i$ and $j$. $a^p_{ij}$ is the connectivity of node $i$ with respect to all the other nodes at a given order of the adjacency matrix $A^p$. $s^p$ is the attenuation factor where $s \in$ [0,1]. $\alpha_p(i)$ and $\alpha_p(j)$ are the PageRank centralities of node $i$ and node $j$, respectively.  $d$ is the damping parameter (set to 0.85 in this study). $v_j$ refers to an eigenvector of the adjacency matrix $A$, associated with its eigenvalue $\lambda_j$.} 
\label{tab_classical_centrality}
 \end{center}
\end{table}

\begin{table}[ht!]
 \begin{center}
   \begin{tabular}{p{6cm}|p{6cm}}
   \hline\hline
   Centrality measure description & Definition \\
   \hline
   \textbf{Community Hub-Bridge} \cite{ghalmane2019immunization}:  based on weighting the intra-community links by the node's community size and the inter-community links by the node's number of neighboring communities  & $
\beta_{CHB}(i) = |c_k| \times k_i^{intra} + |NNC_i| \times k_i^{inter}
$   \\

   \textbf{Participation Coefficient} \cite{guimera2005functional}: based on the heterogeneity of a node's links, where the more external links a node has, the higher its centrality  &  $
\beta_{PC}(i) = 1 - \sum_{c=1}^{N_c} 
\left(
\frac{k_{i,c}}{k_i}
\right)^2
$ \\
   
    \textbf{Community‑based Mediator} \cite{tulu2018identifying}: based on the entropy of a node's intra-community and inter-community links   & $
\beta_{CBM}(i) = H_i \times \frac{k_i}{\sum_{i=1}^{N} k_i}
$

  \\
    \textbf{Comm Centrality} \cite{gupta2016centrality}: based on weighting the intra-community and inter-community links by the proportion of external links and prioritizes bridges  &   

$
\beta_{Comm}(i) = 
(1 + \mu_{c_k}) \times \chi +  (1 - \mu_{c_k})  \times  \varphi^2
$  \\
   
     \textbf{Modularity Vitality} \cite{modvitality}: a signed community-aware centrality based on the modularity change a node causes after its removal from the network  & 
$\beta_{MV}(i) =  M(G_i) - M(G) $\\

   \textbf{Community-based Centrality} \cite{zhao2015community}: based on weighting the intra-community and inter-community links by the size of their belonging communities &  
   $
\beta_{CBC}(i) =  \sum_{c=1}^{N_c} k_{i,c} 
\left(
\frac{n_c}{N}
\right)
$   \\

    \textbf{K-shell with Community} \cite{luo2016identifying}: based on the k-shell hierarchical decomposition of the local network (formed by intra-community links) and the global network (formed by inter-community links) &  
$
\beta_{ks}(i) = \delta \times \beta^{intra}(i) + (1- \delta) \times \beta^{inter}(i) 
$
   \\
 
   \hline
   \end{tabular}
\caption{Definitions of community-aware centrality measures ($\beta(i)$). $c_k$ is the $k$-th community. $k_i^{intra}$ and $k_i^{inter}$ represent the intra-community and inter-community links of a node. $N_c$ is the total number of communities. $k_{i,c}$ is the number of links node $i$ has in a given community $c$.  $k_i$ is the total degree of node $i$. $N$ is the total number of nodes. $H_i = [-\sum \rho_i^{intra} log(\rho_i^{intra})] + [- \sum \rho_i^{inter} log(\rho_i^{inter})]$ is the entropy of node $i$ based on its $\rho^{intra}$ and $\rho^{inter}$ which represent the density of the communities a node links to. $\chi = \frac{k_i^{intra}}{max_{(j \in c)}k_j^{intra}} \times R$ and $\varphi = \frac{k_i^{inter}}{max_{(j \in c)}k_j^{inter}} \times R$. $\mu_{c_k}$ is the proportion of inter-community links over the total community links in community $c_k$. $R$ is a constant to scale intra-community and inter-community values to the same range. $M$ is the modularity of a network and $M(G_i)$ is the modularity of the network after the removal of node $i$. $n_c$ is the number of nodes in community $c$. $\beta^{intra}(i)$ and $\beta^{inter}(i)$ represent the k-shell value of node $i$ by only considering intra-community links and inter-community links, respectively. $\delta$ is set to 0.5 in this study.
} 
\label{tab_communityaware_centrality}
 \end{center}
\end{table}

\begin{table}[ht!]
 \begin{center}
   \begin{tabular}{p{4cm}|p{8cm}}
   \hline\hline
    Domain & Network's name and number\\
\hline
\textbf{Animal networks} & Dolphins (1), Reptiles (2) \\
\textbf{Biological networks} & Budapest Connectome (3),  Blumenau Drug (4), E. coli Transcription (5),  Human Protein (6), Interactome Vidal (7), Kegg Metabolic (8),  Malaria Genes (9), Mouse Visual Cortex (10), Yeast Collins (11), Yeast Protein (12) \\
\textbf{Collaboration networks} & DBLP (13), AstroPh (14), C.S. PhD (15), GrQc (16), NetSci (17), New Zealand Collaboration (18) \\
\textbf{Offline social networks} &  Adolescent health (19), Jazz (20), Zachary Karate Club (21), Madrid Train Bombings (22) \\
\textbf{Infrastructural networks}& EU Airlines (23), EuroRoad (24), Internet Autonomous Systems (25), Internet Topology Cogentco (26), London Transport  (27), U.S. Power Grid  (28), U.S. Airports  (29), U.S. States (30) \\
\textbf{Actor networks} & Game of Thrones (31), Les Misérables (32), Marvel Partnerships (33), Movie Galaxies (34) \\
\textbf{Miscellaneous networks}& 911AllWords (35), Bible Nouns (36), Board of Directors (37), DNC Emails (38), Football (39), Polbooks (40) \\
\textbf{Online social networks}  & DeezerEU (41), Ego Facebook (42), Facebook Friends (43), Facebook Organizations (44), Caltech (45), Facebook Politician Pages (46), Hamsterster (47), PGP (48), Princeton (49), Retweets Copenhagen (50) 
\\
   \hline
   \end{tabular}
\caption{The fifty real-world networks used in this study divided into eight different domains. All network data can be obtained from the cited resources \cite{nr, icon, latora2017complex, netz, kunegis2014handbook}.
} 
\label{tab_data}
 \end{center}
\end{table}

\section*{Correlation Analysis}

\begin{figure}[t]
\begin{center}
\includegraphics[width=12cm, height=8cm]{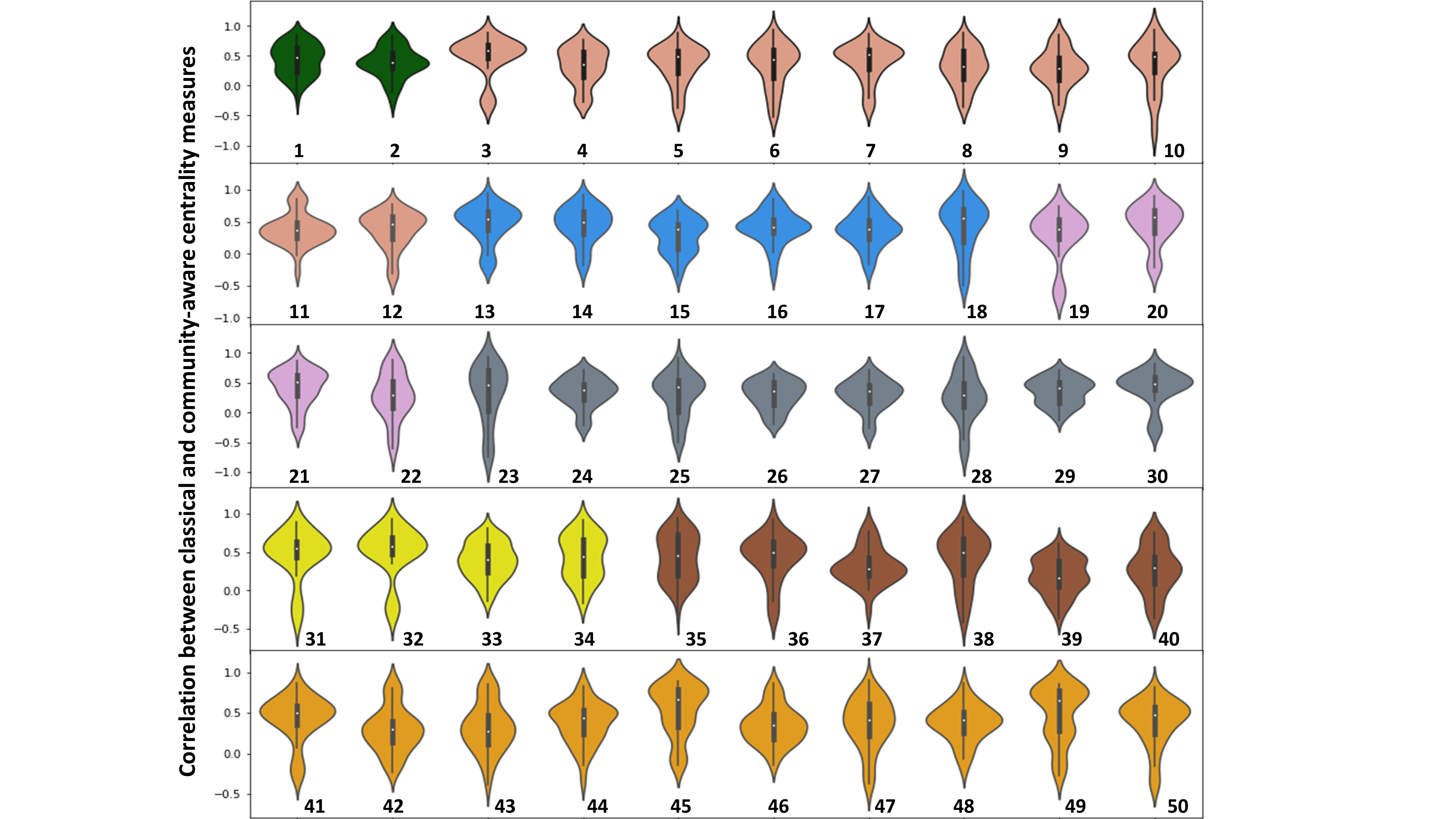}
 \caption{Distribution of the Kendall's Tau correlation between classical and community-aware centrality measures for each network. Colors represent the network's domain. Animal networks are green. Biological networks are Pink. Collaboration networks are blue. Offline social networks are violet. Infrastructural networks are grey. Actor networks are yellow. Miscellaneous networks are brown. Online social networks are orange.} 
 \label{FigureViolinPlotCentralityCorrDistribution}
\end{center}
\end{figure}


The first investigation concerns how classical and community-aware centrality measures correlate for a given network. So, for each of the fifty networks, the Kendall’s Tau correlation is computed for all possible combinations between the ten classical ($\alpha_i$) and seven community-aware centrality measures ($\beta_j$). Figure \ref{FigureViolinPlotCentralityCorrDistribution} shows the distributions of the correlation values for each network. There is no consistency of the distribution for networks from the same domain. Indeed, their distributions can be quite different. For example, although EU Airlines (23) and  EuroRoad (24) belong to the infrastructural networks domain (grey color), EU Airlines (23) has a wide distribution while EuroRoad (24) is much narrow.  One can notice that most networks exhibit a unimodal distribution. Yet, bimodal distributions are also seen, such as in the networks Movie Galaxies (34), 911AllWords (35), and Football (39). Whatever the network considered, the most frequent value of the distribution lies around 0.5. The average median of all the distributions is 0.43$\pm$0.1. The average interquartile range is 0.37$\pm$0.1. Finally, the average mean of the distribution for all networks is 0.37$\pm$0.07. In other words, most of the classical and community-aware centrality measures tend to exhibit medium to low correlation values. Yet, few high correlation values are also observed.



\begin{figure}[t]
\begin{center}
\includegraphics[width=1\textwidth,  height=3cm, keepaspectratio]{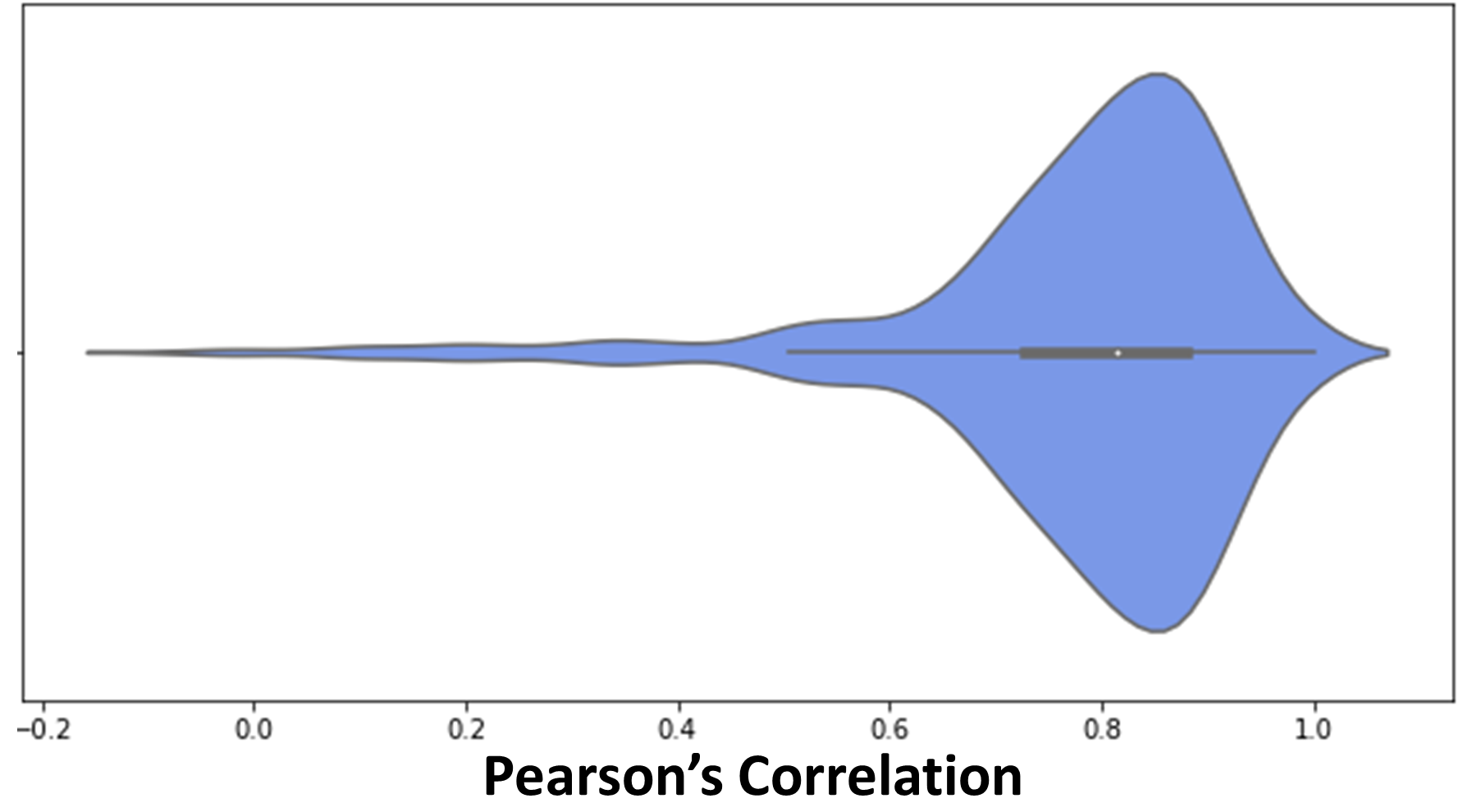}
 \caption{Distribution of Pearson's correlation for the heatmaps of the Kendall's Tau correlation between classical and community-aware centrality of all networks.} 
 \label{FigureViolinPlot50x50}
\end{center}
\end{figure}

To check the consistency of Kendall's Tau correlation values for the various pairs of community-aware and classical centralities across networks, we proceed as follows. Each network is represented by a sample made of thirty-five correlation pair values. The Pearson correlation values between the samples two-by-two are then computed to quantify the two networks' statistical proximity. Figure \ref{FigureViolinPlot50x50} illustrates its distribution. Globally, results across networks are well-correlated. Indeed, the Pearson correlation values range from 0.6 and 1. More precisely, their mean value is equal to 0.80, and their median is 0.82. Note that 911AllWords, Football, and to a lesser extent, Ego Facebook deviate from the general trend. That is the reason why the distribution has a fat left tail. Hence, one can conclude that the correlation of classical and community-aware centrality measures across networks is rather consistent.

\begin{figure}[t]
\begin{center}
\includegraphics[width=1\textwidth,  height=3.7cm, keepaspectratio]{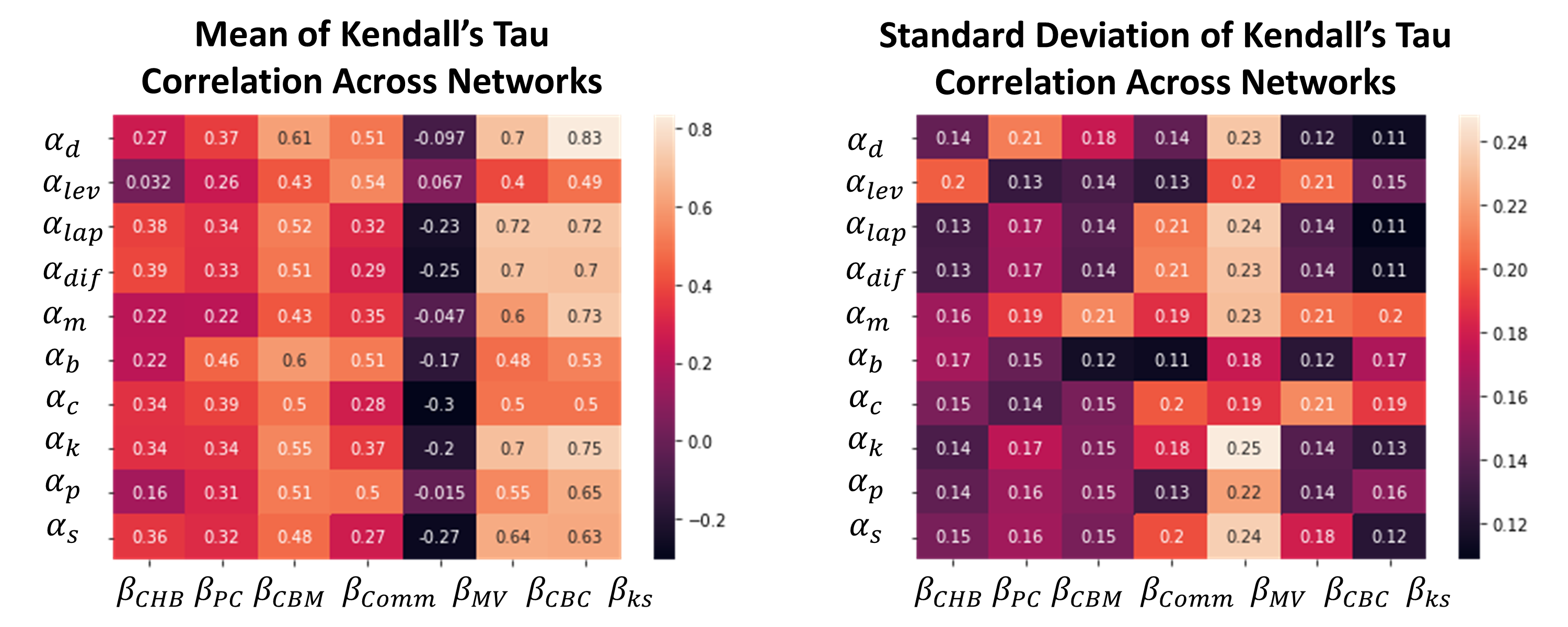}
 \caption{Mean and standard deviation of the Kendall's Tau correlation for each classical and community-aware centrality measures pair ($\alpha_i$, $\beta_j$) across the fifty networks.} 
 \label{FigureMeanStdvOfCombinationsAcrossNetworks}
\end{center}
\end{figure}



Finally, having checked that Kendall's Tau correlation values are consistent across networks, we calculate the mean and standard deviation for each combination ($\alpha_i$, $\beta_j$) across the fifty networks. It allows studying if community-aware centrality measures behave differently. Results reported in figure \ref{FigureMeanStdvOfCombinationsAcrossNetworks} show that the various community-aware centrality measures' correlation patterns are very different. Modularity Vitality ($\beta_{MV}$) is the only community-aware centrality measure exhibiting a negative correlation with classical centrality measures. Furthermore, its mean standard deviation value is high.  As it is a signed community-aware centrality measure, this result is not unexpected. 
The remaining community-aware centrality measures can be ranked according to their correlation values. Community Hub-bridge ($\beta_{CHB}$) and Participation Coefficient ($\beta_{PC}$) tend to show low positive mean correlation with all classical centrality measures ($\leq$ 0.4) except for ($\alpha_b$, $\beta_{PC}$) amounting to 0.46. Their subsequent mean standard deviation is generally close to 0.15. Comm Centrality ($\beta_{Comm}$) has a minimum mean correlation of 0.27 and a maximum mean correlation of 0.54. The standard deviation of $\beta_{Comm}$ ranges from 0.11 to 0.21. Next comes Community-based Mediator ($\beta_{CBM}$), where the mean correlation is between 0.43 and 0.6. Its mean standard deviation is near 0.15 for all combinations except for ($\alpha_m$, $\beta_{CBM}$) amounting to 0.21. Finally, Community-based Centrality ($\beta_{CBC}$) and K-shell with Community ($\beta_{ks}$) exhibit a higher correlation with classical centrality measures than the other community-aware centrality measures. Indeed, the mean correlation may even reach 0.83 as a maximum ($\alpha_d$, $\beta_{ks}$). Their standard deviation is in the range of 0.14 and 0.21. These results corroborate the observation of high values of the correlation in each network's distribution reported in figure \ref{FigureViolinPlotCentralityCorrDistribution}. Indeed, these values correspond to $\beta_{CBC}$ and $\beta_{ks}$.

\section*{Network topology analysis}

\begin{figure}[t!]
\begin{center}
\includegraphics[width=1\textwidth,  height=14cm, keepaspectratio]{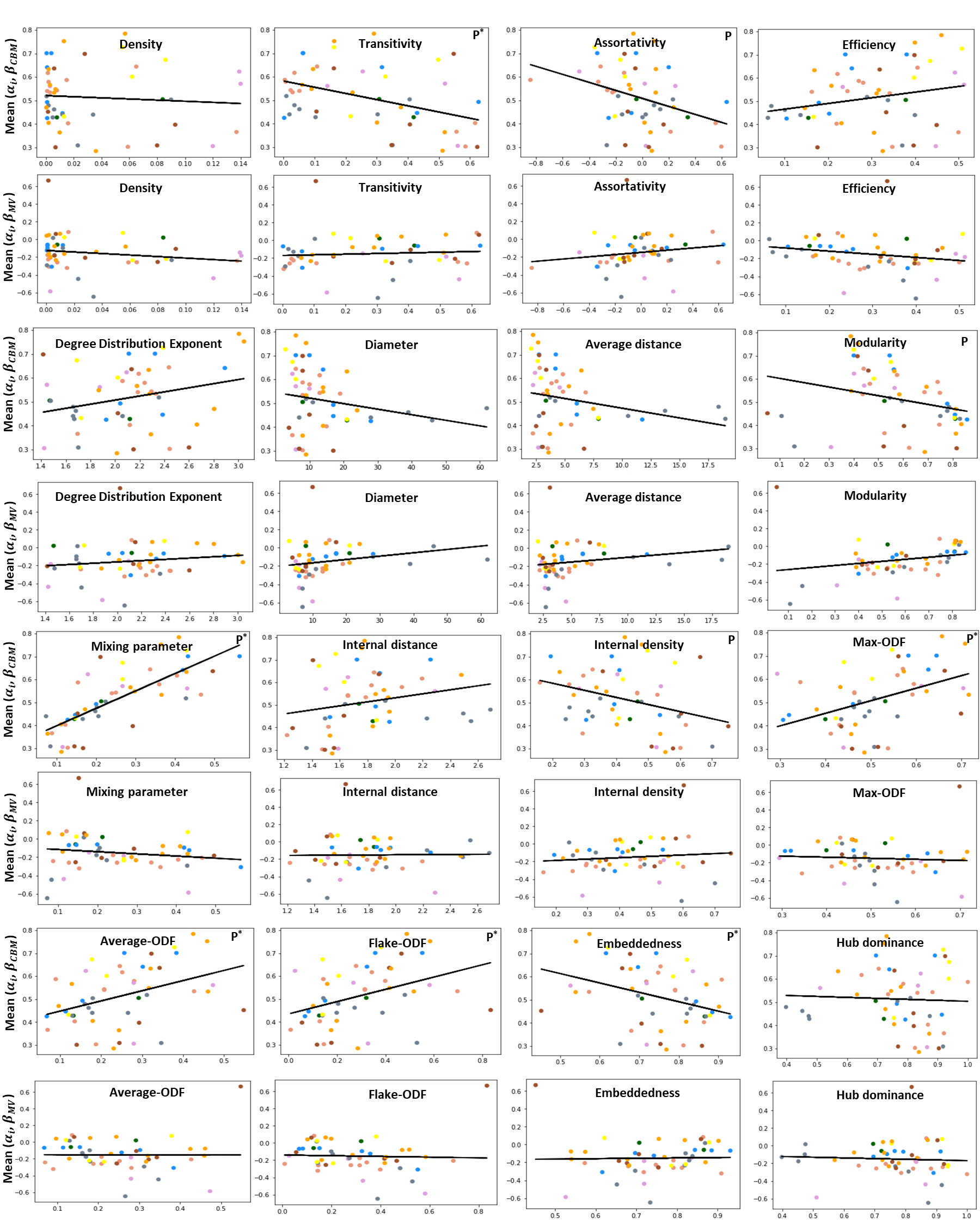}
 \caption{Relationship of the mean of the correlation between the community-aware centralities ``Community-based Mediator ($\beta_{CBM}$)" and ``Modularity Vitality ($\beta_{MV}$)" combined with all classical centrality measures as a function of the topological properties of real-world networks. The line is fitted by linear regression using ordinary least squares. ``P" indicates $p\leq0.05$. ``P" and * indicate $p\leq0.01$. The colors of the data points represent the network's domain.}
  \label{LinearRegression}
\end{center}
\end{figure}

Correlation values between classical and community-aware centrality measures of each network are further processed. For a given network, each community-aware centrality measure is reduced to the mean value of the Kendall's Tau correlation values computed for the ten classical centrality measures. Simple linear regression is performed to investigate the relationship with various topological properties of the networks. The average correlation values are the dependent variables, while the topological properties are the independent variables. The macroscopic features used are Density, Transitivity, Assortativity, Average distance, Diameter, Efficiency, and the Degree distribution exponent. The mesoscopic features used are Modularity, Mixing parameter, Internal distance, Internal density, Max-ODF, Average-ODF, Flake-ODF, Embeddedness, and Hub dominance. If the $p$-value is below 0.05, the dependent and independent variables' relationship is considered statistically significant. Figure \ref{LinearRegression} presents the two extreme cases of statistical dependency between the mean and topological features. The premier case concerns Community-based Mediator (The mean value shows significant linear relationships with nine topological features). The last case is for Modularity Vitality (the mean value shows no meaningful linear relationship with any topological property).  The remaining figures and linear regression parameters estimate for each community-aware centrality measure are provided in the supplementary materials.

Regarding macroscopic topological properties, we observe three situations. Network characteristics exhibit a significant linear relationship with the mean of either three, two, or none community-aware centrality. In that sense, transitivity and efficiency are the most influential macroscopic topological features \cite{orman2013empirical}. They show significant relationships with the mean of three different community-aware centrality measures. Then come density, assortativity, diameter, and average distance that affect two community-aware centrality measures. Finally, the degree distribution exponent is the only topological feature among the macroscopic features that do not show any significant relationship. Transitivity has a significant negative association with the mean of Community-Based Mediator ($\beta_{CBM}$) and Participation Coefficient ($\beta_{PC}$). Indeed, increasing transitivity leads to more triangles in the network. As Community-Based Mediator is based on the entropy of the intra-community and inter-community links of a node, transitivity may increase the difference between the two, resulting in a lower correlation. As the Participation Coefficient also exploits the margin of the proportion of the inter-community and intra-community links, it behaves similarly. One observes a positive association with transitivity for Community-based Centrality ($\beta_{CBC}$). If the whole network forms a single community, $\beta_{CBC}$ reduces to degree centrality \cite{zhao2015community}. Consequently, the correlation between $\beta_{CBC}$ and classical measures tend to increase as transitivity increases. Efficiency has a significant positive association on Comm Centrality ($\beta_{Comm}$), Community-based Centrality ($\beta_{CBC}$), and K-shell with Community ($\beta_{ks}$). An increase in efficiency means that the average shortest path distance in a network is getting smaller. In other words, the network is more efficient when nodes are closely connected. Therefore, community-aware centrality measures tend to be more correlated with classical ones. 
Density has a significant positive association with Comm Centrality ($\beta_{Comm}$) and Community-based Centrality ($\beta_{CBC}$). An increase in density means more links between nodes. Accordingly, $\beta_{Comm}$ and $\beta_{CBC}$ get more analogous to classical centrality measures. Assortativity has a significant negative association with the mean of Community-Based Mediator ($\beta_{CBM}$) and Participation Coefficient ($\beta_{PC}$). An increase in assortativity means that there are more interactions between peers in the networks. It may also increase the margin of difference between intra-community and inter-community links. Assortative networks tend to form communities with ``similar" degree nodes. Consequently, intra-community and inter-community link densities may further differ from one community to another. Hence, a lower correlation between $\beta_{CBM}$/$\beta_{PC}$ and classical centrality measures is observed. Diameter and average distance have both a significant negative association with the mean of Community-based Centrality ($\beta_{CBC}$) and K-shell with Community ($\beta_{ks}$). An increase in both measures means that nodes are more distant from each other. These two community-aware centrality measures are the most sensitive to distance-related measures.


Regarding the mesoscopic topological features, one can distinguish two cases. The mixing parameter, modularity, and Max-ODF are statistically linearly related with the mean of three community-aware centrality measures. Linear dependence exists with the mean of two community-aware centrality measures for the remaining features.
The mixing parameter has a significant positive association with the mean on Community Hub-Bridge ($\beta_{CHB}$), Participation Coefficient ($\beta_{PC}$), and Community-based Mediator ($\beta_{CBM}$). An increase in the mixing parameter translates into a weaker community structure. As a result, these community-aware centrality measures tend to extract similar information compared to classical centrality measures. Modularity has a significant negative association with the mean on Community-based Mediator ($\beta_{CBM}$), Community-based Centrality ($\beta_{CBC}$), and K-shell with Community ($\beta_{ks}$). An increase in modularity means that communities are tightly connected. As a result, these measures extract different information than classical centrality measures when the network is highly modular. Max-ODF has a significant positive association with the mean of Community-based Mediator ($\beta_{CBM}$), Community-based Centrality ($\beta_{CBC}$), and K-shell with Community ($\beta_{ks}$). Based on the nodes with the highest inter-community links in their community, its increase leads to more connections between highly connected nodes in different communities, weakening the community structure. Therefore, correlation of $\beta_{CBM}$, $\beta_{CBC}$, and $\beta_{ks}$ with classical centrality measures increases. Internal distance shows a significant positive linear relationship with the mean of Participation Coefficient ($\beta_{PC}$) and a negative one with the mean of Community-based Centrality ($\beta_{CBC}$). As $\beta_{PC}$ exploits the heterogeneity between intra-community and inter-community links of a node, an increase in internal decreases the margin between intra-community and inter-community links. Consequently, the correlation between $\beta_{PC}$ and classical centrality measures increases. The opposite effect occurs with $\beta_{CBC}$.
Internal density has a negative influence on the mean of Community-based Mediator ($\beta_{CBM}$) and Participation Coefficient ($\beta_{PC}$). An increase in internal density means that communities are condensed with inner connections. As $\beta_{CBM}$ and $\beta_{PC}$ exploit the margin of difference of a node's intra-community links to its inter-community links, both will favor an increase in internal density. Average-ODF has a significant positive relationship with the mean of Community-based Mediator ($\beta_{CBM}$) and K-shell with Community ($\beta_{ks}$). Since it is based on the proportion of inter-community links, the weaker the community structure, the higher the correlation with classical centrality measures. Similarly, Flake-ODF has a similar positive linear relationship with the mean of $\beta_{CBM}$ and $\beta_{ks}$. Indeed, it is another way of quantifying the strength of the community structure. Embeddedness has a negative relationship with the mean of Community-based Mediator ($\beta_{CBM}$) and K-shell with Community ($\beta_{ks}$). Indeed, based on the proportion of intra-community links, it is the opposite of Average-ODF. Finally, hub dominance has a significant positive relationship with the mean of Community-based Centrality ($\beta_{CBC}$) and K-shell with Community ($\beta_{ks}$). A higher hub dominance means fewer tightly connected communities. As a result, $\beta_{CBC}$ behaves closer to degree centrality, and the correlation of $\beta_{CBC}$ with classical centrality measures increases. Concerning $\beta_{ks}$, higher hub dominance induces more similar intra-community and inter-community links and higher correlation with classical centrality measures.

\section*{Conclusion}

This study investigates the relationship between classical and community-aware centrality measures. Results show that the Kendall's Tau correlation between classical and community-aware centrality measures is generally medium to low. Second, the correlation patterns are pretty consistent across networks. Moreover, the community-aware centrality measures can be classified into four groups according to the correlation pattern with classical centrality measures. More specifically, Modularity Vitality shows a low negative correlation. Low positive correlation characterizes Community Hub-Bridge and Participation Coefficient. A positive medium correlation is observed for Comm Centrality and Community-based Mediator. Finally, Community-based Centrality and K-shell with Community show a high positive correlation. Transitivity and efficiency are the most influential macroscopic features while the mixing parameter, modularity, and Max-ODF are the predominant mesoscopic features. The results of this study pave the way for the development of effective community-aware centrality measures. Indeed, it demonstrates that integrating knowledge about the network community structure brings a new perspective of node influence.

%
%
%

\bibliographystyle{splncs04}
\bibliography{biblio}

%





\end{document}